\documentclass{main}
\usepackage{ulem,bm,amsmath,amssymb,braket,subfigure}

\usepackage{textcomp}
\usepackage{fancyhdr}
\usepackage{lastpage}
\def\be{\begin{equation}}
\def\ee{\end{equation}}
\def\bea{\begin{eqnarray}}
\def\eea{\end{eqnarray}}

\newcommand{\md}{\mathrm{d}}
\graphicspath{{figures/}}
\DeclareGraphicsExtensions{.pdf,.png,.jpg,.eps}

\fancypagestyle{firstpagefooter}{%
  \fancyfoot[L]{  \textcopyright \footnotesize This work is an open access article under a Creative Commons Attribution 4.0 International License (https://creativecommons.org/licenses/by/4.0/).}
  \fancyfoot[C]{}  % **Red: This removes the page number from the first page footer**
   
}

\begin{document}

 \title{\Large Azimuthal Modulations in Photon–Induced Processes}

\author{\underline{Ya-jin Zhou}\footnote{Speaker, email: zhouyj@sdu.edu.cn}}

\address{
Key Laboratory of Particle Physics and Particle Irradiation (MOE), Institute of Frontier and Interdisciplinary Science, Shandong University, Qingdao, Shandong 266237, China
}

\maketitle\abstracts{
We review recent theoretical developments and experimental measurements of azimuthal modulations in photon-induced processes, covering both ultra-peripheral heavy-ion collisions (UPCs) and $e^+e^-$ colliders. The azimuthal asymmetries $\cos(n\phi)$ ($n=1,2,3,4$) serve as precision diagnostics that probe the linear polarization of coherent photons, final-state soft radiation effects, and quantum interference phenomena at the femtometer scale. In UPCs, we discuss dilepton production, diffractive dijet and vector meson production, where theoretical predictions show excellent agreement with STAR, ALICE, and CMS measurements. The unique double-slit interference effect in nucleus-nucleus collisions plays a crucial role in describing experimental observations. At $e^+e^-$ colliders, $\gamma\gamma\to\pi\pi$ azimuthal asymmetries enable the direct extraction of helicity amplitude phases, with important implications for hadronic light-by-light scattering and the muon anomalous magnetic moment. Azimuthal modulations establish a powerful tool for multi-dimensional nuclear imaging and precision QED/QCD tests.
}

%\footnotesize DOI: \url{https://doi.org/xx.yyyyy/nnnnnnnn}


\section{Introduction and theoretical origin}

Ultra-peripheral collisions (UPCs) provide a clean laboratory for photon-induced reactions at high energies, where hadronic interactions are suppressed and electromagnetic fields dominate. The equivalent photon approximation (EPA), pioneered by Fermi \cite{Fermi:1924tc} and by Weizsäcker–Williams \cite{vonWeizsacker:1934nji,Williams:1934ad}, reformulates these fields as fluxes of quasi-real photons, enabling photon–photon and photon–target reactions.

A second historical thread concerns polarization. The linear polarization of high-energy radiation was already recognized in the classical literature on bremsstrahlung. Sommerfeld analyzed polarization patterns at low electron energies \cite{Sommerfeld:1931}, while later work by May and Wick extended the study to the relativistic regime \cite{Wick:1951,MayWick:1951,May:1951}. Although these early investigations predate modern QED and QCD language, their basic insight—namely, that polarization leaves characteristic angular fingerprints—remains central to current UPC studies.

In fact the linearly polarized photon has a very simple picture.  A static electric field in vacuum is isotropic. However, for an ultrarelativistic charged ion the field is Lorentz-contracted into a thin pancake perpendicular to the beam direction. Over a small patch of this transverse plane, the electric field points approximately radially, so the local field looks like a plane wave with a definite linear polarization. If one integrates over the entire transverse plane and over all photon transverse momenta, the preferred directions average out for an unpolarized nucleus and the net polarization vanishes. But typical measurements select restricted kinematics—most notably a limited range of photon transverse momentum $k_\perp$—so the residual linear polarization becomes observable. This qualitative picture dovetails naturally with modern transverse-momentum–dependent (TMD) factorization. In 2001, Mulders and Rodrigues formulated a TMD framework that includes a linearly polarized gluon distribution inside an unpolarized hadron \cite{Mulders:2000sh}, 
\begin{eqnarray} 
 \int \frac{2dy^- d^2y_\perp}{xP^+(2\pi)^3} e^{ik \cdot y} \langle P
|  F_{+}^i(0) U_{[0,y]} F_{+}^j(y) U'_{[y,0]}  |P \rangle \big|_{y^+=0}
%\\ \nonumber &&~~~~~
=
\delta_\perp^{i j} f_1(x,\bm{k}_\perp^2)+ \left (\frac{2k_\perp^i
k_\perp^j}{\bm{k}_\perp^2}-\delta_\perp^{ij} \right )
 h_1^{\perp }(x,\bm{k}_\perp^2) ,  \label{eq:gTMDs}
\end{eqnarray}
where $f_1(x,\bm{k}_\perp^2)$ and $h_1^\perp(x,\bm{k}_\perp^2)$ denote the unpolarized and linearly polarized gluon TMDs, with Wilson lines $U_{[0,y]}$ and $U'_{[y,0]}$ ensuring gauge invariance. By analogy, photons from unpolarized charges can be described similarly, except that no Wilson lines are needed since photons are colorless and gauge invariant. At small $x$ — relevant for high-energy nuclei and protons — both gluons and photons are predicted to be strongly linearly polarized~\cite{Metz:2011wb,Li:2019yzy}. This follows because, for a highly relativistic hadron, the Lorentz boost enhances the $A^+$ component, making $F_{+}^i \approx k_\perp^i A^+$ dominant at small $x$. In this limit, comparing Eq.~\eqref{eq:gTMDs} shows $f_1(x,\bm{k}_\perp^2) = h_1^{\perp}(x,\bm{k}_\perp^2)$—indicating maximal linear polarization for dipole-type gluons or photons.

 How are these polarization effects observed? Historically, the asymmetry was defined as ratio of the cross section parallel to the polarization and the cross section perpendicular to it~\cite{Olsen:1959zz}. Today it is standard to analyze the full azimuthal distribution and to extract Fourier moments. Linear polarization of the initial state generates a $\cos(2\phi)$ modulation of the differential cross section, with $\phi$ the angle between a chosen reference axis and the transverse momentum that balances the final state. %Higher harmonics can appear as well: $\cos(4\phi)$ may arise from multiple-scattering effects, from higher-order radiation, or from intrinsic elliptic correlations of the target’s gluon Wigner distribution in diffractive channels.

Azimuthal structure, however, does not come exclusively from initial-state polarization. Final-state radiation (FSR) of soft photons or gluons can also generate anisotropies  \cite{Hatta:2020bgy,Hatta:2021jcd}, especially at relatively large transverse momentum of the pair or jet system. The physical picture is straightforward: if a lepton or parton emits a soft quantum, it recoils, and the recoil shifts the configuration away from the symmetric back-to-back geometry. The resulting distribution acquires a calculable angular dependence with $\cos(2\phi)$, $\cos(4\phi)$, and higher harmonics. These effects can be treated systematically within soft-collinear effective theory (SCET), which resums large logarithms and accommodates mass effects when needed. In practice, measured azimuthal patterns in UPCs receive contributions from both sources—initial linear polarization at small $q_T$ and FSR at moderate $q_T$—so a realistic phenomenology must account for their interplay.

%Finally, UPCs also enable a third class of angular phenomena rooted in coherence and interference at the Fermi (nuclear) scale. When the photon wavelength is comparable to the size of the emitter or when multiple spatially separated sources contribute coherently—as in two colliding nuclei—double-slit–like interference can imprint modulations in azimuth and in qT. In diffractive channels this coherence couples to the spatial distribution of gluons in the nucleus, and specific harmonics become sensitive to geometric anisotropies, offering tomographic information about small-x nuclear structure.

In the remainder of this contribution we apply these general ideas to several flagship processes: dilepton production \cite{Li:2019yzy,Li:2019sin,Shao:2022stc,Shao:2023zge} and LbL scattering in UPCs \cite{Jia:2024hen}, diffractive dijet production where both ISR and FSR matter \cite{Shao:2024nor}, diffractive vector meson production with its sensitivity to gluon geometry and Wigner distributions \cite{Xing:2020hwh,Hagiwara:2020juc,Hagiwara:2021xkf,Brandenburg:2022jgr}, and, finally, light-meson pair production in $e^+e^-$ collisions \cite{Jia:2024xzx}, where the cleaner environment allows a controlled exploration of polarization-induced modulations using an interdiscipline of TMD factorization, chiral perturbation theory (ChPT), and dispersion relations. Across these examples, azimuthal harmonics serve as precision diagnostics that go beyond total cross sections, turning angle-dependent patterns into quantitative probes of polarization, soft radiation, and the spatial dynamics of QCD at small x.

\section{Azimuthal modulations in different photon-induced processes}

We now turn to specific photon-induced reactions that we have studied, and where azimuthal modulations have been measured or are within experimental reach. Each process highlights different aspects of the mechanisms discussed above: initial-state linear polarization, final-state radiation, and nuclear-structure effects.

\subsection{Photon-photon collision processes in UPC}

\subsubsection{Dilepton production in UPC}
The production of di-lepton pairs ($e^+e^-$ or $\mu^+\mu^-$) in UPCs provides the cleanest arena for testing the polarized-EPA framework. This process was the first where azimuthal modulations from linearly polarized photons were systematically predicted and subsequently confirmed by experiment.

Within the joint impact parameter and TMD framework, the di-lepton production via photon-photon fusion can be described by $\gamma_1(x_1 P+\tilde{k}_{1\perp})+\gamma_2(x_2 \bar{P}+\tilde{k}_{2\perp}) \rightarrow l^+(p_1)+ l^-(p_2)$, where the leptons are produced in a nearly back-to-back configuration with $q_\perp=|\bm{q}_\perp| \ll P_\perp=|\bm{P}_\perp|$. The joint $\bm{b}_\perp$ and $\bm{q}_\perp$ dependent di-lepton production cross section at the lowest order of QED can be written as \cite{Li:2019sin},
\begin{eqnarray}
 \frac{d \sigma_{0}}{d^{2} \bm{p}_{1\perp} \bm{p}_{2\perp} d y_{1} d y_{2} d^{2} \bm{b}_{\perp}} = \frac{2\alpha_e^2}{(2\pi)^2Q^4}
\left [ \mathcal{A}+ \mathcal{B} \cos 2\phi+\mathcal{C} \cos 4\phi \right ]
\end{eqnarray}\label{born}
with
\begin{eqnarray}
\mathcal{A}\! & = & \!{\cal \int}[{d\cal K}_\perp]\frac{1}{\left(P_{\perp}^{2}+m^{2}\right)^{2}} \Bigl[-2m^{4}\cos\left(\phi_{\bm{k}_{1\perp}}+\phi_{\bm{k}_{1\perp}'}-\phi_{\bm{k}_{2\perp}}-\phi_{\bm{k}_{2\perp}'}\right)+m^{2}\left(Q^{2}-2m^{2}\right) \nonumber \\ &&\times 
\cos\left(\phi_{\bm{k}_{1\perp}}-\phi_{\bm{k}_{1\perp}'}-\phi_{\bm{k}_{2\perp}}+\phi_{\bm{k}_{2\perp}'}\right)%\nonumber\\ &  & 
 +P_{\perp}^{2}\left(Q^{2}-2P_{\perp}^{2}\right)\cos\left(\phi_{\bm{k}_{1\perp}}-\phi_{\bm{k}_{1\perp}'}+\phi_{\bm{k}_{2\perp}}-\phi_{\bm{k}_{2\perp}'}\right)\Bigr],  \\
\mathcal{B} \! & = & \!{\cal \int}[{d\cal K}_\perp]\frac{8m^{2}P_{\perp}^{2}}{ \left(P_{\perp}^{2}+m^{2}\right)^{2}} \cos\left(\phi_{\bm{k}_{1\perp}}-\phi_{\bm{k}_{2\perp}}\right)\cos\left(\phi_{\bm{k}_{1\perp}'}+\phi_{\bm{k}_{2\perp}'}-2\phi\right),\\
\mathcal{C} \! & = & \!{\cal \int}[{d\cal K}_\perp]\frac{-2P_{\perp}^{4}}{\left(P_{\perp}^{2}+m^{2}\right)^{2}}\cos\left(\phi_{\bm{k}_{1\perp}}+\phi_{\bm{k}_{1\perp}'}+\phi_{\bm{k}_{2\perp}}+\phi_{\bm{k}_{2\perp}'}-4\phi\right), 
\end{eqnarray}
where the shorthand notation represents
\begin{eqnarray}\nonumber
{\cal \int}[d{\cal K}_\perp ]&\equiv& \int d^{2}\bm{k}_{1\perp}d^{2}\bm{k}_{2\perp}d^{2}\bm{k}_{1\perp}'d^{2}\bm{k}_{2\perp}'e^{i(\bm{k}_{1\perp}-\bm{k}_{1\perp}')\cdot \bm{b}_{\perp}} %\nonumber \\&\times& 
\delta^{2}(\bm{k}_{1\perp}+\bm{k}_{2\perp}-\bm{q}_{\perp}) \delta^{2}(\bm{k}_{1\perp}'+\bm{k}_{2\perp}'-\bm{q}_{\perp}) \\ && \times 
\mathcal{F}(x_1,\bm{k}_{1\perp}^{2})\mathcal{F}(x_2, \bm{k}_{2\perp}^{2})\mathcal{F}(x_1, \bm{k}_{1\perp}'^{2})\mathcal{F}(x_2, \bm{k}_{2\perp}'^{2}). 
\end{eqnarray}
The function $\mathcal{F}(x_{1}, \bm{k}_{\perp}^{2})$ is related to the photon TMD via  
\begin{eqnarray}
    |\mathcal{F}(x_{1}, \bm{k}_{\perp}^{2})|^2=x_1 f_1^\gamma(x_1,\bm{k}_{\perp}^2)=
\frac{Z^2 \alpha_e}{\pi^2} \bm{k}_{\perp}^2
\left [ \frac{F(\bm{k}_{\perp}^2+x^2M_p^2)}{(\bm{k}_{\perp}^2+x^2M_p^2)}\right ]^2,
\label{f1h1}
\end{eqnarray}
where $F$ is the Woods-Saxon form factor
\begin{eqnarray} 
F(\bm k^2)= \int d^3 \bm{r} e^{i\bm k\cdot \bm r} \frac{\rho^0}{1+\exp{\left [(r-R_{\mathrm{WS}})/d\right ]}}.
\end{eqnarray}

Higher-order QED contributions introduce final-state soft photon radiation effects, generating large logarithmic terms $\alpha_e^n {\rm ln}^{2n} \frac{Q^{2}}{m^2}$ that can be resummed using the Collins-Soper formalism \cite{Collins:1984kg}. The resummed cross section reads \cite{Klein:2018fmp,Hatta:2021jcd},
\begin{eqnarray}
  \frac{d\sigma}{d^2 \bm{p}_{1\perp} d^2 \bm{p}_{2\perp} dy_1 dy_2 d^2 \bm{b}_\perp }= \int
  \frac{d^2 \bm{r}_\perp}{(2\pi)^2} e^{i \bm{r}_\perp \cdot \bm{q}_\perp} e^{- \mathrm{Sud}(r_\perp)} \int d^2 \bm{q}_\perp'
  e^{-i \bm{r}_\perp \cdot \bm{q}_\perp'} \frac{d\sigma_{_{\!0}}(\bm q_\perp')}{ d\mathcal{P.S.}}, \label{eq:res1}
\end{eqnarray}
where $d {\cal P.S.}=d^2 \bm{p}_{1\perp} d^2 \bm{p}_{2\perp} dy_1 dy_2 d^2 \bm{b}_\perp $ and the Sudakov factor at one-loop order is
\begin{eqnarray}\label{eq:sud_DL} 
  \mathrm{Sud}(\mu_r,r_\perp)=
\frac{\alpha_e}{\pi} {\rm ln} \frac{Q^2}{m^2}  {\rm ln}\frac{P_\perp^2}{\mu_r^2},
\end{eqnarray}
with $\mu_r=2 e^{-\gamma_E}/r_{\perp}$. The azimuthal asymmetries are defined as
\begin{eqnarray}
\langle \cos(n\phi) \rangle =\frac{ \int \frac{d \sigma}{d {\cal P.S.}} \cos (n\phi) \ d {\cal P.S.} }
{\int \frac{d \sigma}{d {\cal P.S.}}  d {\cal P.S.}}.
\end{eqnarray}

Early theoretical predictions for $\langle \cos(4\phi) \rangle$ in di-electron production \cite{Li:2019yzy,Li:2019sin} were promptly verified by RHIC STAR collaboration \cite{Adam:2019mby}. As shown in Table \ref{tab:v4}, the measured $\langle \cos(4\phi) \rangle$ for $\gamma \gamma \to e^+e^-$ aligns excellently with QED calculations.
\begin{table}[hbpt]
    \centering
    \begin{tabular}{|c|c|c|}
    \hline
         & Measured $|2\langle \cos(4\phi) \rangle|$ & QED calculated $2\langle \cos(4\phi) \rangle$\\
    \hline
     UPC & 16.8\%$\pm$2.5\% & $-$16.5\% \\
    \hline
60\%-80\%& 27\%$\pm$6\%     & $-$34.5\%  \\
    \hline
    \end{tabular}
     \caption{Theoretical and experimental results for $\cos(4\phi)$ asymmetry in di-electron UPC production in Au+Au collisions at RHIC ($\sqrt{s} = 200$ GeV). Kinematic cuts: $Q \in [0.45, 0.76]$ GeV, $y_{1,2} \in [-1, 1]$, $q_\perp < 0.1$ GeV, $P_\perp > 0.2$ GeV.}
    \label{tab:v4}
    \label{tab:v4}
\end{table}

An important extension of this framework addresses the detailed interplay between initial-state polarization and final-state radiation effects in muon pair production \cite{Shao:2022stc}. The $\cos(2\phi)$ modulation receives contributions from both the linearly polarized photon TMD and soft photon radiation, which can be systematically separated through kinematic constraints. At small $q_\perp$, the $\cos(2\phi)$ asymmetry is dominated by initial-state polarization effects, while at moderate $q_\perp$ values, final-state radiation contributions become increasingly important. When complete muon mass corrections are included in the resummation formula, the azimuthal asymmetries receive sizable corrections, particularly at relatively large pair transverse momentum $q_\perp$ at RHIC energies. The mass-dependent Sudakov factor modifies the resummation structure, and numerical studies reveal that the $\langle \cos(2\phi) \rangle$ and $\langle \cos(4\phi) \rangle$ asymmetries can differ significantly from the massless approximation when $q_\perp \gtrsim 0.1$ GeV. This effect becomes more pronounced at lower invariant masses and demonstrates the importance of retaining full mass dependence in precision studies of UPC dilepton azimuthal correlations.

The complete resummation framework for azimuthal asymmetries in dilepton production has been systematically derived using SCET and standard RG methods \cite{Shao:2023zge}. This approach allows for resummation of lepton mass effects to all orders in perturbation theory. The SCET framework and RG evolution method provide a systematic treatment of the large logarithms that arise from FSR resummed to all orders. The resulting resummation formula provides a unified description across different kinematic regimes, enabling robust predictions for azimuthal asymmetries at both RHIC and LHC energy regions. Our results show that the $q_\perp$-dependent azimuthal asymmetries are relatively insensitive to subleading resummation effects, but the leading single-logarithm contribution is essential for describing the acoplanarity data from ATLAS and CMS.

\subsubsection{Light-by-light process in UPCs}
Two-photon collisions can produce pairs of photons via the pure QED process $\gamma\gamma\to\gamma\gamma$, known as light-by-light (LbL) scattering. This reaction is particularly interesting because it probes QED at loop level and has sensitivity to possible beyond-Standard-Model (BSM) physics, such as axion-like particles or other light weakly coupled states.

In our analysis\cite{Jia:2024xzx}, we calculated the UPC cross section for LbL scattering including azimuthal modulations. The total cross section and basic kinematic distributions agree with predictions from established Monte Carlo generators such as \textsc{SuperChic} \cite{Harland-Lang:2020veo} and \textsc{gamma-UPC} \cite{Shao:2022cly}. However, when comparing with ATLAS data, some kinematic regions show intriguing tensions that warrant further investigation.

Beyond total rates, azimuthal-angle observables offer complementary diagnostics. The $\cos(2\phi)$ modulation probes the linear polarization structure of the incoming photons, while higher harmonics can arise from final-state photon interactions or from interference between direct LbL scattering and related processes. 

Specifically, the differential cross section reads

\begin{eqnarray}\label{eq:dsig_helicity}
\frac{d\sigma}{d^{2}\bm{p}_{1\perp}d^{2}\bm{p}_{2\perp}dy_{1}dy_{2}d^{2}\bm{b}_{\perp}}
%----------------------------
%\nonumber \\
%----------------------------
 & = & \frac{1}{32\pi^{2}Q^{4}}\int d^{2}{\bm k}_{1\perp}d^{2}{\bm k}_{2\perp}\frac{d^{2}{\bm k}_{1\perp}'}{(2\pi)^{2}}
 \delta^{2}({\bm q}_{\perp}-{\bm k}_{1\perp}-{\bm k}_{2\perp})
%----------------------------
% \nonumber \\
%----------------------------
% & \times & 
 e^{i({\bm k}_{1\perp}-{\bm k}_{1\perp}')\cdot {\bm b}_{\perp}}
 \nonumber \\ \nonumber
&& \Big\{ \cos(\phi_{1}-\phi_{2})\cos(\phi_{1}'-\phi_{2}')|M_{++}|^{2}
%----------------------------
% \nonumber \\
%----------------------------
+\cos(\phi_{1}+\phi_{2})\cos(\phi_{1}'+\phi_{2}')|M_{+-}|^{2}
%----------------------------
\label{eq:cross section}
%----------------------------
 \\
%----------------------------
 & &\quad -\cos(\phi_{1}+\phi_{2})\cos(\phi_{1}'-\phi_{2}')M_{++}M_{+-}^{*}
%----------------------------
 \nonumber \\ \nonumber
%----------------------------
 & & \quad -\cos(\phi_{1}-\phi_{2})\cos(\phi_{1}'+\phi_{2}')M_{+-}M_{++}^{*} \Big\}
%----------------------------
\\
%----------------------------
 && \,\times\,   {\cal F}(x_{1},{\bm k}_{1\perp}^{2})\,{\cal F}^{*}(x_{1},{\bm k}_{1\perp}'^{2})\,{\cal F}(x_{2},{\bm k}_{2\perp}^{2})\,{\cal F}^{*}(x_{2},{\bm k}_{2\perp}'^{2}),
%----------------------------
 %\nonumber
%----------------------------
\end{eqnarray}
where $\phi_{1,2}'$ denote the azimuthal angles between ${\bm k}_{1,2\bot}'$ and ${\bm P}_{\bot}$, and $M_{\lambda_{1},\lambda_{2}}$ is related with the
helicity amplitude of $\gamma(x_1 P,\lambda_1)\gamma(x_2\overline{P},\lambda_{2})\to\gamma(p_{1})+\gamma(p_{2})$.
The occurrence of $M_{\lambda_{1},\lambda_{2}}M_{\lambda_{1}^{\prime},\lambda_{2}^{\prime}}^{*}$ represents the short-hand
notation for
%----------------------------
\begin{eqnarray}
%----------------------------
M_{\lambda_{1},\lambda_{2}}M_{\lambda_{1}^{\prime},\lambda_{2}^{\prime}}^{*} \equiv
 \sum_{\lambda_3,\lambda_4} M_{\lambda_{1},\lambda_{2},\lambda_3,\lambda_4} M^*_{\lambda_{1}^{\prime},\lambda_{2}^{\prime},\lambda_3,\lambda_4}.
%----------------------------
\end{eqnarray}

These angular observables serve multiple purposes: they provide handles to distinguish SM contributions from BSM scenarios, since new light states would modify both total rates and angular patterns in characteristic ways; they also help constrain theoretical uncertainties in the loop-level QED calculation itself.

While current LbL measurements at the LHC focus primarily on total cross sections and basic kinematic distributions, future high-statistics datasets should enable extraction of azimuthal coefficients. This would open a new window for precision QED tests and BSM searches in a theoretically clean channel.

\subsection{Photon-nuclear collision processes in UPCs}
In ultra-peripheral heavy-ion collisions, coherent photons from one nucleus can interact with the other nucleus, leading to photon-nucleus reactions. The main theoretical tools for treating high-energy photon-nucleus reaction processes are the color dipole model and the color glass condensate (CGC) effective theory. Within these theoretical frameworks, the photon first splits into a quark-antiquark pair, forming a color dipole moment. The color dipole then interacts with the nuclear matter in the heavy ion, either forming two jets — referred to as diffractive dijet production — or recombining to form a vector meson. In this section, we will study both types of diffractive production.

\subsubsection{Diffractive di-jet production in UPCs}

Diffractive di-jet production in photon-nucleus collisions provides a unique window into the three-dimensional structure of gluonic matter in nuclei. In the correlation limit where the two jets are nearly back-to-back ($P_\perp \gg q_\perp$), the transverse momentum imbalance $q_\perp$ encodes information about gluon Wigner distributions, which describe the simultaneous distribution of gluons in both position and momentum space. The azimuthal modulation $\cos(2\phi)$, where $\phi$ is the angle between $\bm q_\perp$ and $\bm P_\perp=(\bm k_{1\perp}-\bm k_{2\perp})/2$, serves as a direct probe of the elliptic gluon Wigner distribution.

Recent theoretical developments have revealed that semi-inclusive diffractive di-jet production, characterized by a tri-jet configuration where a semi-hard gluon is emitted along the beam direction, significantly dominates over exclusive di-jet production. This dominance arises because the color octet $q\bar{q}$ pair in the tri-jet setup can scatter strongly via gluon exchange, circumventing the color transparency suppression that typically suppresses exclusive production. We revisit the azimuthal asymmetry analysis in this framework, focusing on the impact of both the initial-state and final state soft gluon radiations~\cite{Shao:2024nor}.

\begin{figure}
    \centering
    \includegraphics[width=0.9\linewidth]{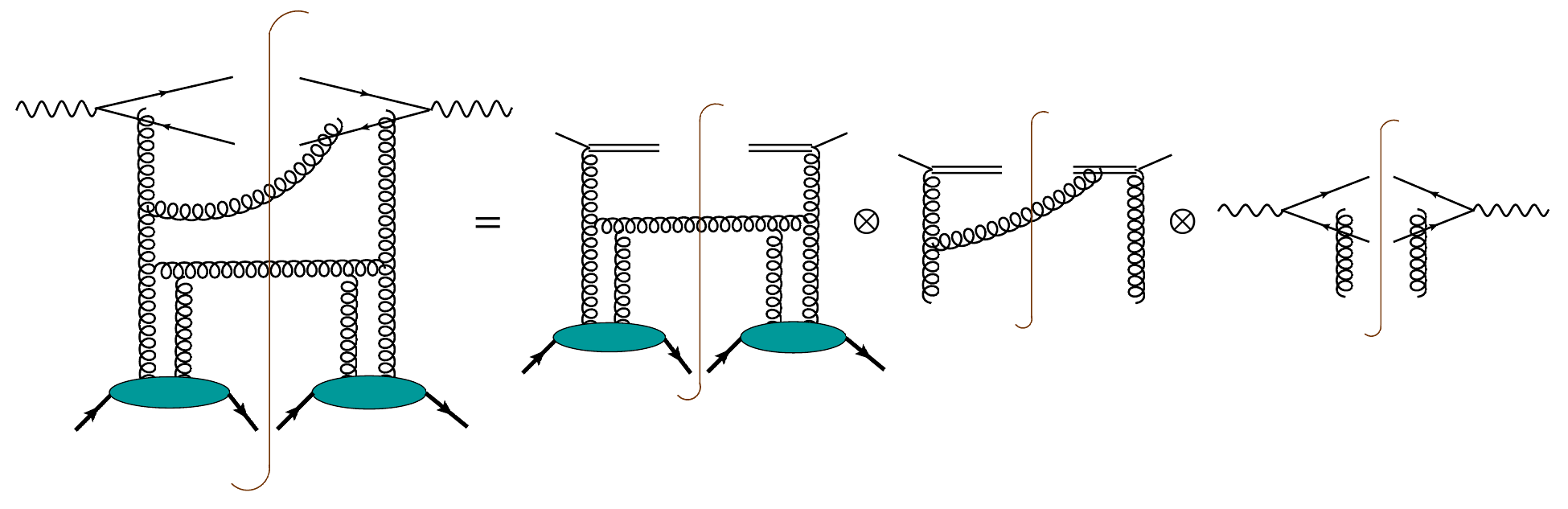}
    \caption{An illustration of the factorization scheme. The double line represents the gauge link.}
    \label{fig:dijet_factorization}
\end{figure}

The process can be described within the CGC formalism, where the photon splits into a $q\bar{q}$ pair, and one of the partons emits a gluon. The three-parton system then scatters elastically off the nuclear target. The factorized cross section in the SCET framework reads
\begin{eqnarray}
\frac{\md \sigma}{\md y_1 \, \md y_2 \, \md^2 \bm P_\perp \, \md^2 \bm q_\perp }=&\,\sigma_0  x_\gamma f_\gamma(x_\gamma) H_{\gamma^* g}(P_\perp,R,\mu) \int \md^2 \bm k_\perp \md^2 \bm \lambda_\perp \delta^{(2)}(\bm q_\perp - \bm k_\perp - \bm \lambda_\perp) \nonumber \\
&\times  S(\bm \lambda_\perp,R,\mu) \int \frac{\md x_{\mathbb{P}}}{x_{\mathbb{P}}} x_g G_{\mathbb{P}}^{\rm unsub}(x_g, x_{\mathbb{P}}, k_\perp, \mu),
\end{eqnarray}
where $G_{\mathbb{P}}$ is the diffractive transverse momentum dependent (DTMD) gluon distribution, $S$ is the soft function accounting for radiations from initial and final states, and $H_{\gamma^* g}$ is the hard function. The factorization scheme is illustrated in Fig.~\ref{fig:dijet_factorization}.

The azimuthal asymmetry receives contributions from both FSR and ISR. FSR tends to align soft gluons with jet directions, generating a $\cos(2\phi)$ modulation. However, ISR exhibits axisymmetric behavior, which naturally suppresses the azimuthal asymmetry. By performing all-order resummation of both contributions using the Collins-Soper formalism and renormalization group evolution, we find that ISR significantly broadens the $q_\perp$ distribution due to leading double-logarithm enhancement, while FSR contributes at the single-logarithm level.

We compared our numerical results for UPCs at LHC energies with CMS measurements. %in Fig.~\ref{fig:dijet_cms}.
The result show that the inclusion of ISR effects provides a more accurate description of the $q_\perp$ distribution. The azimuthal asymmetry is suppressed at low $q_\perp$ when ISR is included, consistent with the axisymmetric nature of ISR. While our calculations capture the overall trends, some discrepancies remain, suggesting that additional mechanisms may be at play. The study of these normalized observables, which are free from overall cross-section normalization uncertainties, offers a unique opportunity to explore the production mechanisms of hard dijets in diffractive processes and to probe gluon saturation effects in nuclei.

\subsubsection{Diffractive vector meson production in UPCs}
Diffractive vector meson production in UPCs provides a powerful probe of gluon structure in nuclei. At leading order in the Eikonal approximation, the polarization state of coherent photons is fully transmitted to the final-state vector meson, making the decay products anisotropic rather than isotropic. 
We have systematically studied azimuthal asymmetries in diffractive vector meson production, including the $\cos(2\phi)$ asymmetry in $\rho^0$ production~\cite{Xing:2020hwh}, the $\cos\phi$ and $\cos(3\phi)$ asymmetries~\cite{Hagiwara:2020juc}, the $\cos(4\phi)$ asymmetry~\cite{Hagiwara:2021xkf}, as well as the $\cos(2\phi)$ effect in $J/\psi$ production~\cite{Brandenburg:2022jgr}.

\begin{figure}[hbt]
    \centering
    \includegraphics[width=0.4\linewidth]{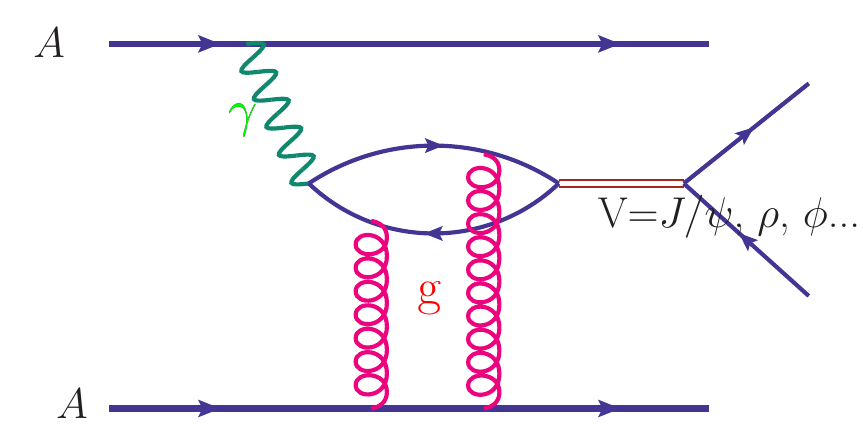}
    \caption{Schematic diagram for diffractive vector meson production in UPC. }
    \label{fig:vm_diagram}
\end{figure}

The process is described within the color dipole model, where the quasi-real photon is viewed as a $q\bar{q}$ color dipole that undergoes multiple scattering with the CGC gluon field from the other heavy ion, as illustrated in Fig.~\ref{fig:vm_diagram}. The Eikonal approximation plays a crucial role: quarks and antiquarks maintain their relative transverse positions and spin states unchanged, so the polarization state of the color dipole is preserved through multiple scattering and transmitted to the final-state vector meson. The amplitude for photon-nucleus reactions can be expressed as a convolution in coordinate space of the photon light-cone wave function (describing the photon fluctuation into a $q\bar{q}$ pair), the vector meson wave function (describing the meson fluctuation into a $q\bar{q}$ pair), and the dipole scattering amplitude:
\begin{eqnarray}
{\cal A}(\Delta_\perp)=i \!\int \! d^2
b_\perp e^{i \Delta_\perp \cdot b_\perp} \!\int \! \frac{d^2
r_\perp}{4\pi} \! \int_0^1 \!\! dz \  \Psi^{\gamma\rightarrow q\bar q}
 (r_\perp,z,\epsilon_{\perp}^{\gamma})N(r_\perp,b_\perp)
\Psi^{ V\rightarrow q \bar q *} (r_\perp,z,\epsilon_{\perp}^{V}),
\end{eqnarray}
where $-\Delta_\perp$ is the nuclear recoil transverse momentum, $z$ is the light-cone momentum fraction carried by the quark, and $\epsilon_{\perp}^{\gamma}$ and $\epsilon_{\perp}^{V}$ are the polarization vectors of the quasi-real photon and final-state vector meson, respectively. $N(r_\perp,b_\perp)$ is the scattering amplitude for a color dipole of size $r_\perp$ with the target nucleus at impact parameter $b_\perp$.

A unique feature of vector meson production in heavy-ion collisions is the double-slit interference effect. When the vector meson mass is much larger than the inverse nuclear radius, the meson can be approximately localized at a transverse position $b_\perp$. The production amplitude at a given $b_\perp$ is proportional to the electromagnetic potential from one colliding nucleus and the gluon matter density from the other target nucleus. Since both nuclei can alternately serve as the photon source and gluon target, and we cannot distinguish between these two production scenarios, we must add the two amplitudes before squaring, giving rise to interference terms. This interference effect, analogous to Young's double-slit experiment,  is crucial for correctly describing the experimental observations~\cite{Xing:2020hwh}.

\begin{figure}
    \centering
    \includegraphics[width=0.4\linewidth]{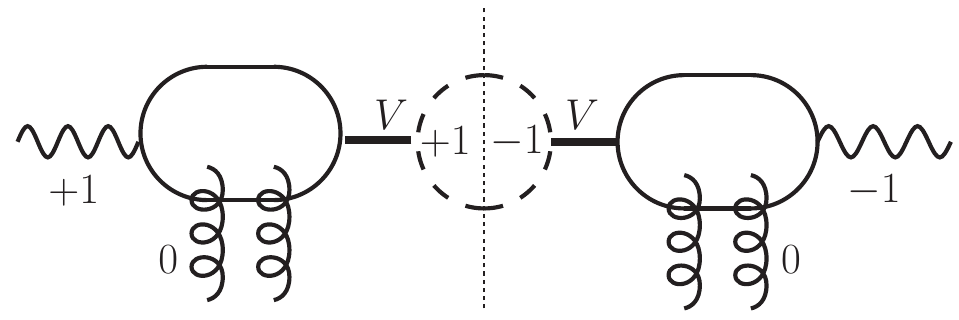}
    \includegraphics[width=0.4\linewidth]{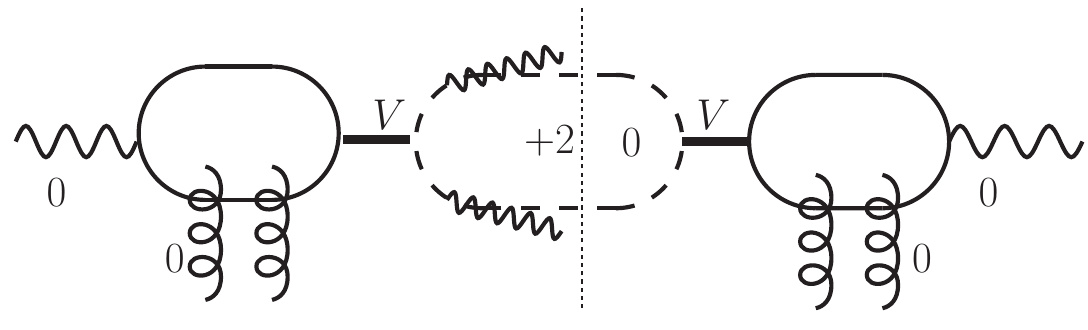}
    \caption{Illustration diagrams for $\cos2\phi$ azimuthal asymmetry}
    \label{fig:cos2phi}
\end{figure}

For $\rho^0$ production, the differential cross section for $\rho^0 \to \pi^+\pi^-$ decay reads:
\begin{eqnarray}
  && \!\!\!\!\!\!\!\!\!\!\!\!
   \frac{d \sigma_{\rho \rightarrow \pi \pi}}{d^2 p_{1\perp} d^2 p_{2\perp} dy_1 dy_2 d^2 \tilde b_{\perp} }  =\frac{1 }{2(2\pi)^7}  \frac{P_\perp^2}{(Q^2-M_\rho^2)^2+M_\rho^2 \Gamma_\rho^2}  f_{\rho \pi \pi}^2 
    \nonumber \\ &&\!\!\!\!\!\!\!\!\times
    \int d^2 \Delta_\perp d^2k_\perp d^2 k_\perp'
    \delta^2(k_\perp+\Delta_\perp-q_\perp)
     (\hat P_\perp \! \cdot \hat k_\perp )(\hat P_\perp\! \cdot \hat k_\perp' )
     \nonumber \\ && \!\!\!\!\!\!\!\!\times \!
    \left \{ \int \!\! d^2   b_\perp e^{i \tilde b_\perp \cdot
    (k_\perp'\!\!-k_\perp)} \left [ T_A(b_\perp) {\cal
    A}_{in}(x_2,\Delta_\perp) {\cal A}_{in}^*(x_2,\Delta_\perp') {\cal
    F}(x_1,k_\perp){\cal F}(x_1,k_\perp')\!
      + \!( A \!\leftrightarrow \! B)
    \right ] \right .\
     \nonumber \\ &&
    + \!\!\left [  e^{i \tilde b_\perp \cdot (k_\perp'\!\!-k_\perp)}
    {\cal A}_{co}(x_2,\Delta_\perp) {\cal A}_{co}^*(x_2, \Delta_\perp')
    {\cal F}(x_1,k_\perp){\cal F}(x_1,k_\perp')
     \right ]
      \nonumber \\ &&
    + \!\!\left [  e^{i \tilde b_\perp \cdot (\Delta_\perp'\!\!-\Delta_\perp)}
    {\cal A}_{co}(x_1,\Delta_\perp) {\cal A}_{co}^*(x_1, \Delta_\perp')
    {\cal F}(x_2,k_\perp){\cal F}(x_2,k_\perp')
     \right ]
    \nonumber \\  &&+ \!\!
      \left [ e^{i \tilde b_\perp \cdot (\Delta_\perp'\!-k_\perp)}
     {\cal A}_{co}(x_2,\Delta_\perp) {\cal A}_{co}^*(x_1, \Delta_\perp'){\cal F}(x_1,k_\perp){\cal F}(x_2,k_\perp')
     \right ]
        \nonumber \\  &&+ \!\!\!\!
     \left .\
     \left [ e^{i \tilde b_\perp \cdot (k_\perp'\!-\Delta_\perp)}
     {\cal A}_{co}(x_1,\Delta_\perp) {\cal A}_{co}^*(x_2, \Delta_\perp'){\cal F}(x_2,k_\perp){\cal F}(x_1,k_\perp')
     \right ]
      \right \}\!
       \label{fcs}
     \end{eqnarray}
where $y_1$ and $y_2$ are the rapidities of $\pi^+$ and $\pi^-$, respectively. The term $(\hat P_\perp \! \cdot \hat k_\perp )(\hat P_\perp\! \cdot \hat k_\perp' )$ contains the azimuthal asymmetry information of the decay products. The photon polarization state is transferred to $\rho^0$, and their spin correlation converts to the correlation between $\rho^0$'s transverse polarization and transverse momentum. Due to angular momentum conservation, the $\rho^0\to\pi^+\pi^-$ decay amplitude ${\cal M} \propto e^{i \lambda \phi}$, where $\phi$ is the azimuthal angle of $\pi$ relative to $\rho^0$ and $\lambda$ represents the $\rho^0$ spin state. The interference between amplitudes with $\lambda=+1$ and $\lambda=-1$ gives rise to the $\cos(2\phi)$ azimuthal asymmetry. We show the illustration diagram in Fig.\ref{fig:cos2phi} (left).  The differential cross section for $J/\psi$ production is similar. 
\begin{figure}
    \centering
    \includegraphics[width=0.4\linewidth]{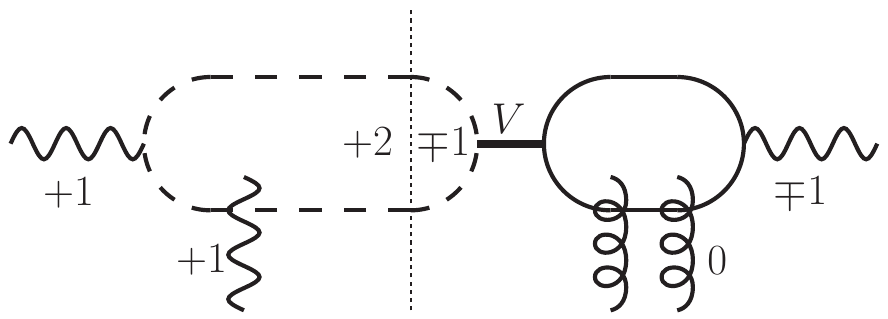}
    \caption{Illustration diagram for $\cos3\phi$/$\cos\phi$ azimuthal asymmetry}
    %, $\braket{+2|\mp 1} \sim \cos 3\phi/\cos \phi$.}
    \label{fig:cosphi}
\end{figure}
\begin{figure}
    \centering
    \includegraphics[width=0.4\linewidth]{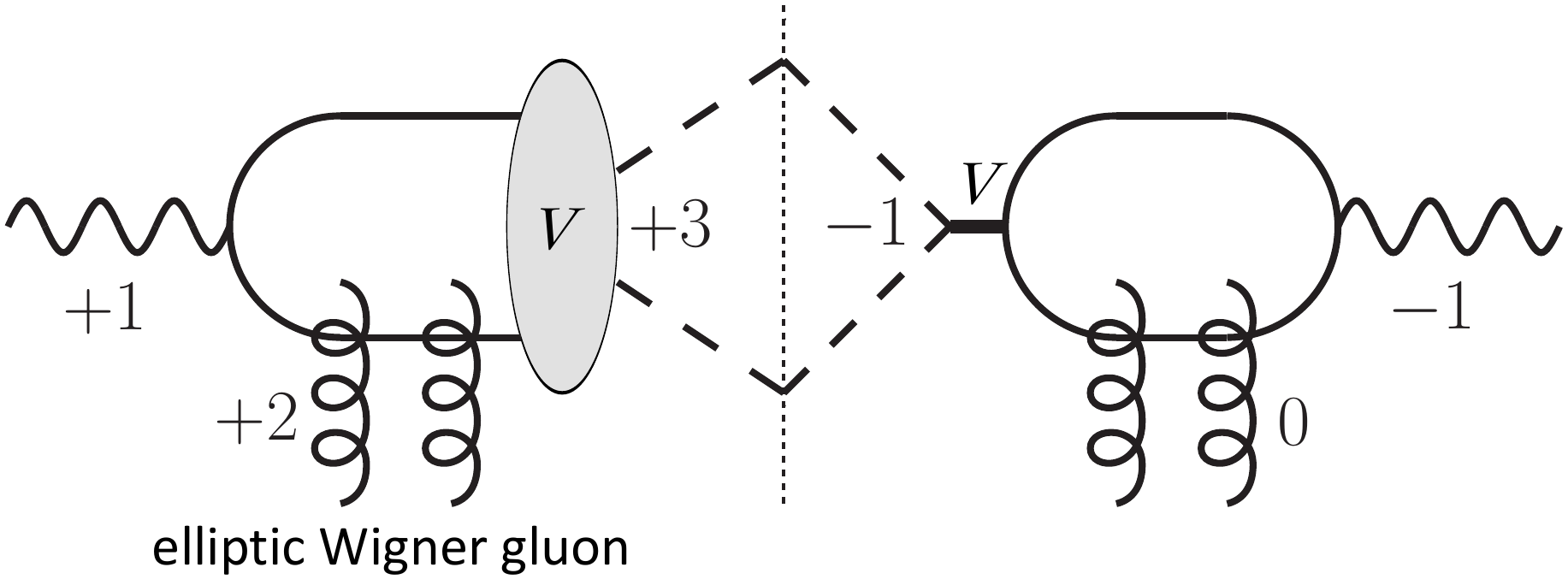}
    \includegraphics[width=0.4\linewidth]{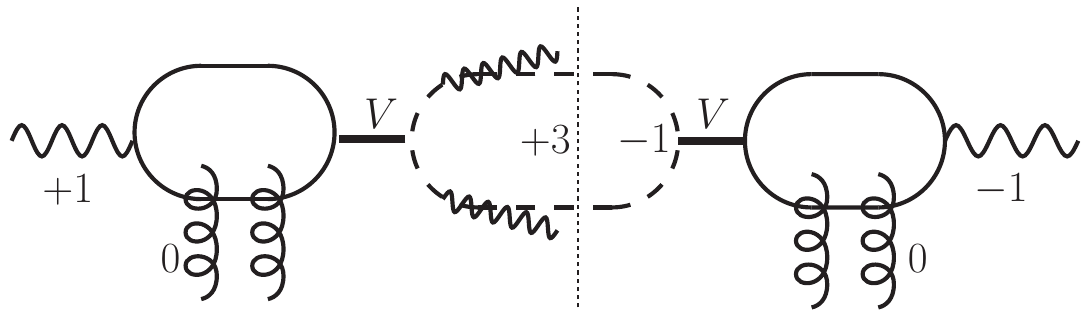}
    \caption{Illustration diagram for $\cos 4\phi$ azimuthal asymmetry}
    \label{fig:cos4phi}
\end{figure}

Another origin of azimuthal modulations is the soft photon radiation effect, as we have explained in the introduction and also in the first subsection of this section. 
The azimuthal asymmetry induced by recoiling leptons can be equivalently understood as angular momentum transfer, as we illustrate in our schematic diagrams. For example, the $\cos(2\phi)$ asymmetry arising from soft photon radiation is demonstrated in Fig.~\ref{fig:cos2phi} (right).

In the $\pi^+\pi^-$ invariant mass spectrum, besides the $\rho^0$ Breit-Wigner resonance, there is also direct electromagnetic production via photon-photon fusion, which becomes comparable to the $\rho^0$ decay at very low transverse momentum. When both photons are in the $+1$ or $-1$ helicity state, interference between direct $\pi^+\pi^-$ production and $\rho^0$ decay leads to $\cos\phi$ or $\cos3\phi$ azimuthal asymmetries~\cite{Hagiwara:2020juc}. These odd harmonics arise from interference between amplitudes with different orbital angular momentum: $\cos\phi$ from $\braket{+2|+1}$ pattern and $\cos3\phi$ from the $\braket{+2|-1}$ pattern. We show this in Fig.\ref{fig:cosphi}.

The $\cos(4\phi)$ asymmetry can arise from two mechanisms~\cite{Hagiwara:2021xkf}. First, when two Wigner gluons are exchanged, they effectively carry 2 units of orbital angular momentum. Together with an additional 1 unit from the linearly polarized photon, this transfers 3 units to the vector meson, and interference with $\rho^0$ decay production ($\lambda=-1$) leads to the $\cos(4\phi)$ asymmetry, as shown in Fig.~\ref{fig:cos4phi} (left). Second, final-state soft photon radiation from the $\pi$ meson pair can be treated as contributing 2 units of angular momentum, which, combined with the linearly polarized photon contribution, gives rise to the $\cos(4\phi)$ asymmetry, as shown in Fig.\ref{fig:cos4phi} (right).

Numerical calculations for $\rho^0$ production show excellent agreement with STAR data for both the unpolarized cross section and the $\cos(2\phi)$ asymmetry~\cite{Xing:2020hwh}. The $\cos(2\phi)$ asymmetry is almost entirely generated in coherent scattering, even though both coherent and incoherent contributions affect the unpolarized cross section. The STAR Collaboration performed measurements in Au+Au UPCs at $\sqrt{s_{NN}}=200$ GeV \cite{STAR:2023SciAdv}. Their results for azimuthal modulations are consistent with our theoretical calculations. The ALICE Collaboration measured the $\cos2\phi$ azimuthal asymmetry at $\sqrt{s}=5.02$ TeV, as well as its impact-parameter dependence for the first time~\cite{ALICE:2024ife}, also shows very good ingreement with our predictions. For the $\cos(4\phi)$ asymmetry, theoretical calculations including contributions from both final-state soft photon radiation and the elliptic gluon Wigner distribution can reproduce the observed peak structure, though the magnitude is still somewhat smaller than experimental values from STAR Collaboration.

The $\cos\phi$ and $\cos3\phi$ asymmetries are particularly significant at very low transverse momentum where the direct contribution is comparable to the resonance decay~\cite{Hagiwara:2020juc}. The STAR Collaboration also extracted $\cos\phi$ and $\cos3\phi$ azimuthal asymmetries from the $\rho^0\to\pi^+\pi^-$ decay channel. While the statistical precision is still limited, the measured shapes are consistent with our predictions based on interference between resonant $\rho^0$ production and direct $\pi\pi$ production.

Similar studies have been performed for $J/\psi$ production~\cite{Brandenburg:2022jgr}. The charmed quark mass provides a hard scale, making perturbative calculations more reliable than for $\rho^0$. The $\cos(2\phi)$ asymmetry mechanism is similar: linearly polarized photons produce a color dipole, which scatters with the other nucleus via unpolarized CGC gluons to form $J/\psi$, which then decays to $e^+e^-$ or $\mu^+\mu^-$. The photon polarization is transferred to $J/\psi$ and reflected in the azimuthal anisotropy of the decay products. Theoretical calculations for LHC energies show good agreement with experimental measurements for the unpolarized cross sections, and predict significant $\cos(2\phi)$ asymmetries at low $q_\perp$ due to the double-slit interference effect and photon linear polarization, and at moderate $q_\perp$ due to final-state soft photon radiation.

The study of these spin-dependent observables, which are sensitive to nuclear structure, opens new theoretical and experimental pathways for multi-dimensional imaging of nuclei. The unique double-slit interference effect in nucleus-nucleus collisions, absent in electron-nucleus collisions at the Electron–Ion Collider (EIC), provides complementary information that will enable precision tests of our understanding of gluon structure in nuclei at small $x$.

\subsection{Light meson pair production at $e^+e^-$ colliders}

While UPCs provide a natural laboratory for photon-induced processes, $e^+e^-$ colliders offer complementary advantages. In particular, the process $\gamma\gamma\to M\bar{M}$ (where $M$ is a light meson such as $\pi$, $K$, or $\eta$) can be studied with excellent control over initial-state kinematics and backgrounds.

The key advantage of $e^+e^-$ over UPCs for this channel lies in avoiding strong hadronic backgrounds from $\rho$ meson resonances. In UPCs, $\rho$ production and decay can contaminate $\gamma\gamma\to\pi\pi$ measurements, making it difficult to isolate the pure two-photon fusion signal. At $e^+e^-$ machines, the initial-state photons come from lepton beams with well-defined energies, allowing for cleaner event selection and better control over resonance contributions.

In our analysis\cite{Jia:2024xzx}, we applied TMD factorization to $\gamma\gamma\to\pi\pi$ in the regime where the $\pi\pi$ pair has small transverse momentum relative to the individual pion momenta ($p_{\pi\pi} \ll p_\pi$). This is precisely the kinematic region where TMD descriptions are most reliable and where polarization effects are expected to be largest.

The theoretical framework combines several elements: (1) TMD factorization for the $\gamma\gamma\to q\bar{q}$ hard process with linearly polarized photons, (2) ChPT to describe the transition from the $q\bar{q}$ pair to the final-state pions, and (3) data-driven dispersion relations to constrain the non-perturbative $\pi\pi$ production amplitude. This interdisciplinary approach connects TMD physics, low-energy effective field theory, and dispersive methods in a novel way.

 The differential cross section in the helicity amplitude form can be obtained from \eqref{eq:dsig_helicity} by taking $k_{i\perp }=k_{i\perp }'$ and $\phi_i=\phi_i'$ with $i=1,2$, since impact-parameter dependence is absent here. The most important feature of this framework is that the $\cos(2\phi)$ azimuthal asymmetry provides a direct means to extract the relative phase between the helicity amplitudes $M_{++}$ and $M_{+-}$. This is in sharp contrast with all preceding studies, where the phases of the helicity amplitudes are extracted through indirect methods, {\it e.g.}, by combining dispersive techniques with experimental input. The accurate knowledge of the corresponding partial wave amplitudes provides the key input for the dispersive determination of the hadronic LbL contribution, which comprises a major source of uncertainty in theoretical prediction of muon anomalous magnetic moment.

\begin{figure}[htbp]
%----------------------------
\centering{
\includegraphics[width=0.3\textwidth]{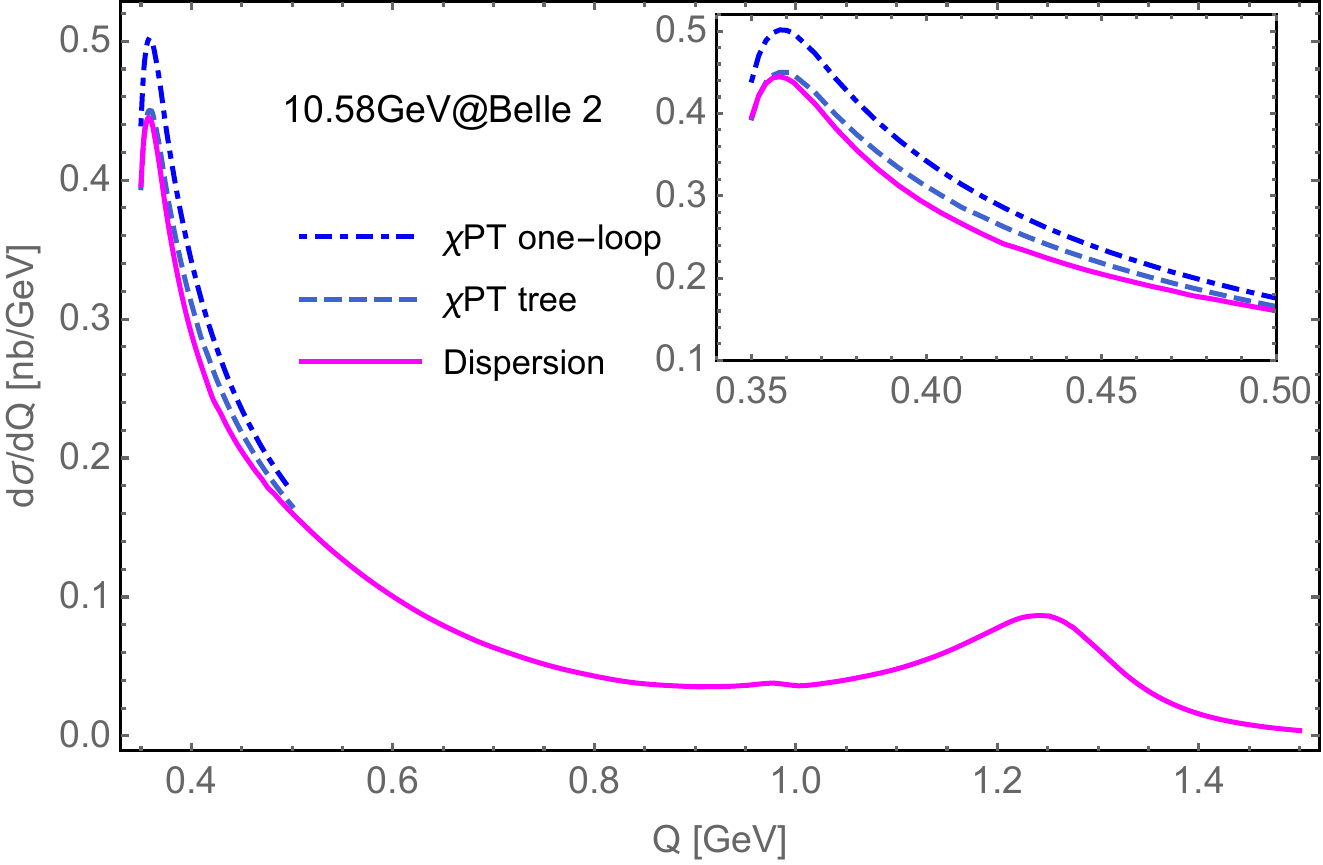}
\includegraphics[width=0.3\textwidth]{asQ_pi+pi-1}
\includegraphics[width=0.3\textwidth]{asQ_pi+pi-1}}
%----------------------------
 \caption{Dipion differential cross section (left), $\langle \cos 2\phi \rangle$ (middle) and $\langle \cos 4\phi \rangle$ (right) asymmetries vs. dipion invariant mass for $e^+e^- \rightarrow \gamma \gamma e^{+}e^{-}\rightarrow \pi^+\pi^- e^{+}e^{-}$ at $\sqrt{s}=10.58$ GeV (Belle 2). Kinematic cuts: $|y_{1,2}| \le 0.38$, $|{\bf P}_\perp|>100$ MeV, $q_\perp \in [0, 50]$ MeV.}
\label{asQ_pi+pi-}
\end{figure}

Our predictions (Fig.\ref{asQ_pi+pi-}) show that the $\cos(2\phi)$ modulation is very large and should be measurable at $e^+e^-$ facilities. This represents the first systematic study of azimuthal modulations for composite light mesons produced via two-photon fusion. The observable is sensitive to both the photon polarization structure and the hadronic matrix elements that connect quarks to pions.

Beyond $\pi\pi$, this framework can be extended to $K\bar{K}$ and $\eta\eta$ production, offering additional handles on chiral dynamics and meson structure. Furthermore, the $\gamma\gamma\to\pi\pi$ process is closely related to the meson-loop contribution to LbL scattering, which enters calculations of the muon anomalous magnetic moment. Understanding azimuthal modulations in $\gamma\gamma\to\pi\pi$ can thus inform both hadronic physics and precision QED tests.

\section{Conclusions and outlook}
Azimuthal-angle modulations provide a unified and sensitive probe of linearly polarized photons/gluons, final-state soft radiation, and interference effects across a wide range of photon-induced processes. In this contribution, we have systematically reviewed theoretical frameworks and experimental measurements of $\cos(n\phi)$ asymmetries ($n=1,2,3,4$) in several key channels.

In UPCs, the production of dilepton pairs via photon-photon fusion offers the cleanest testbed for the polarized-EPA framework, with excellent agreement between theoretical predictions and STAR measurements of $\cos(4\phi)$ asymmetries. The interplay between initial-state polarization and final-state radiation effects has been systematically studied using SCET and RG methods, revealing the importance of complete mass corrections in precision studies. Light-by-light scattering provides complementary probes of QED at loop level and sensitivity to BSM physics through azimuthal observables.

For photon-nucleus collisions in UPCs, diffractive dijet production reveals the impact of initial-state soft gluon radiations on both the $q_\perp$ distribution and azimuthal modulations. Our revisit of this framework shows that ISR effects significantly broaden the $q_\perp$ distribution while suppressing azimuthal asymmetries due to their axisymmetric nature. In diffractive vector meson production, the rich azimuthal structure—including $\cos\phi$, $\cos(2\phi)$, $\cos(3\phi)$, and $\cos(4\phi)$ asymmetries—arises from multiple mechanisms: direct photon polarization transfer, interference between resonant and direct production, elliptic gluon Wigner distributions, and final-state soft photon radiation. The unique double-slit interference effect at the femtometer scale, arising from the ambiguity regarding which nucleus is the photon source, has been confirmed by both theoretical predictions and experimental measurements at RHIC and LHC.

At $e^+e^-$ colliders, the $\gamma\gamma\to\pi\pi$ process offers a superior environment for studying azimuthal modulations due to the absence of strong hadronic backgrounds. The most significant advance is that $\cos(2\phi)$ azimuthal asymmetries enable the direct extraction of the relative phase between helicity amplitudes $M_{++}$ and $M_{+-}$, in contrast to all preceding indirect methods. This direct phase extraction has profound implications for understanding $C$-even resonance structures and for the dispersive determination of hadronic light-by-light contributions to the muon anomalous magnetic moment.

Looking forward, several directions show particular promise. First, the accumulation of high-statistics datasets at Belle 2 and BESIII should enable precise measurements of azimuthal asymmetries in $\pi\pi$, $K\bar{K}$, and $\omega\omega$ production, opening new windows on meson structure and chiral dynamics. Second, future UPC measurements with improved precision will allow detailed studies of the interplay between initial-state polarization, final-state radiation, and nuclear structure effects. Third, global analyses combining multiple harmonics and different processes should enable simultaneous constraints on TMDs  and Wigner distributions. Fourth, the extension of these frameworks to other channels such as $\gamma\gamma\to p\bar{p}$ and $\gamma\gamma\to \rho\rho$ will further test the universality of polarization effects and explore new physics opportunities.

The study of azimuthal modulations in photon-induced processes has evolved from a theoretical curiosity to a precision tool for nuclear tomography and QCD/QED studies. As experimental capabilities continue to advance, these observables will play an increasingly central role in our understanding of hadronic structure, polarization dynamics, and the interplay between perturbative and non-perturbative QCD effects.

\section*{Acknowledgments}
We thank the organizers of UPC 2025 and our collaborators for their efforts on the researches. This work was supported in part by the National Natural Science Foundation of China (Grants No.~12475084) and Shandong Province Natural Science Foundation (Grants No.~ZR2024MA012). Due to space limits we provide only a concise overview and refer the reader to the cited literature for details.

% Appendix removed for brevity in this short contribution.

\end{document}